\documentclass[aps,prb,
preprint,
superscriptaddress,floats]{revtex4-1}
\usepackage{txfonts}
\usepackage{amssymb}
\usepackage{graphicx}
\usepackage{sidecap}
\usepackage{CJK}
\usepackage{txfonts}
\usepackage{url}
\usepackage{verbatim}
\usepackage[normalem]{ulem}
\usepackage{color}
\definecolor{RED}{rgb}{1,0,0}
\definecolor{BLUE}{rgb}{0,0,1}

\begin{document}

%\title{Extraordinarily broad band  and strong electron-phonon coupling in  Ba$_{0.51}$K$_{0.49}$BiO$_3$ superconductor}
\title{Unveiling the superconducting mechanism of   Ba$_{0.51}$K$_{0.49}$BiO$_3$}

\author{C. H. P. Wen}
\author{H. C. Xu} \email{xuhaichao@fudan.edu.cn}
\author{Q. Yao}
\author{R. Peng}
\author{X. H. Niu}
\affiliation{State Key Laboratory of Surface Physics, Department of Physics, and Laboratory of Advanced Materials, Fudan University, Shanghai 200438, People's Republic of China}
\author{Q. Y. Chen}
\affiliation{Science and Technology on Surface Physics and Chemistry Laboratory, Mianyang 621908, China}
\author{Z. T. Liu}
\author{D. W. Shen}
\affiliation{CAS Center for Excellence in Superconducting Electronics (CENSE), Shanghai 200050, China}
\affiliation{State Key Laboratory of Functional Materials for Informatics, Shanghai Institute of Microsystem and Information Technology (SIMIT), Chinese Academy of Sciences, Shanghai 200050, China}
\author{Q. Song}
\author{X. Lou}
\author{Y. F. Fang}
\author{X. S. Liu}
\author{Y. H. Song}
\affiliation{State Key Laboratory of Surface Physics, Department of Physics, and Laboratory of Advanced Materials, Fudan University, Shanghai 200438, People's Republic of China}
\author{Y. J. Jiao}
\author{T. F. Duan}
\author{H. H. Wen}
\affiliation{National Laboratory of Solid State Microstructures and Department of Physics,  Nanjing University, Nanjing 210093, China}
\affiliation{Collaborative Innovation Center of Advanced Microstructures, Nanjing 210093, China}
\author{P. Dudin}
\affiliation{Diamond Light Source, Harwell Science and Innovation Campus, Didcot OX11 0DE, United Kingdom}
\author{G. Kotliar}
\affiliation{Department of Physics, Rutgers University, Piscataway, New Jersey 08854, U.S.A.}
\author{Z. P. Yin}\email{yinzhiping@bnu.edu.cn}
\affiliation{Department of Physics and Center for Advanced Quantum Studies, Beijing Normal University, Beijing 100875, China}
\author{D. L. Feng}\email{dlfeng@fudan.edu.cn}
\affiliation{State Key Laboratory of Surface Physics, Department of Physics, and Laboratory of Advanced Materials, Fudan University, Shanghai 200438, People's Republic of China}
\affiliation{Collaborative Innovation Center of Advanced Microstructures, Nanjing 210093, China}
\date{\today}

\maketitle

\textbf{Bismuthates were the first family of oxide high-temperature superconductors \cite{BaPb_discover}, exhibiting superconducting transition temperatures ($\mathbf{T_c}$) up to 32~K (Refs.~\onlinecite{discover, discover2}), but the superconducting mechanism remains under debate despite more than 30 years of extensive research.
%However, more than 30 years of continuous research has not settled the fundamental question of their superconducting mechanism.
Our angle-resolved photoemission spectroscopy studies 
on Ba$_{0.51}$K$_{0.49}$BiO$_3$ reveal an unexpectedly 34\% larger bandwidth than in conventional density functional theory calculations. This can be reproduced by calculations that fully account for
long-range Coulomb interactions --- the first direct demonstration of bandwidth expansion due to the Fock exchange term, a long-accepted and yet uncorroborated fundamental effect in many body physics              \cite{ashcroft}. 
Furthermore, we observe an isotropic superconducting gap with $\mathbf{2\Delta_0/k_B T_c=3.51\pm0.05}$, and strong electron-phonon interactions with a coupling constant $\mathbf{\lambda\sim}$ 1.3 $\mathbf{\pm}$ 0.2. These findings solve a long-standing mystery --- Ba$_{0.51}$K$_{0.49}$BiO$_3$ is an extraordinary Bardeen-Cooper-Schrieffer (BCS) superconductor, where long-range Coulomb interactions expand the bandwidth, enhance electron-phonon coupling, and generate the high $\mathbf{T_c}$. Such effects will also be critical for finding new superconductors.
}
% The superconductivity in Ba$_{1-x}$K$_{x}$BiO$_3$ with a transition temperature ($T_c$) as high as 32 K was discovered 30 years ago. However, its superconducting mechanism remains unclear, partially due to the lack of direct measurements of its electronic structure and superconducting gap structure. 	

%Bismuth based  superconductors, BaPb$_x$Bi$_{1-x}$O$_3$ and Ba$_{1-x}$K$_{x}$BiO$_3$,  were  the first  family of oxide high temperature superconductors. Discovered, before the cuprates,  they form in a   perovskite structure with superconducting transition temperature ($T_c$) up to 32~K \cite{BaPb_discover,discover, discover2,isotope2,LAPW_Mattheiss,isotope1,Lattice_Fleming}. However,  more than 30 years of continuous research  has not settled the fundamental question of  the nature of the  superconducting mechanism  in these materials.

\begin{figure}[t]
\centering
\includegraphics[width=120mm]{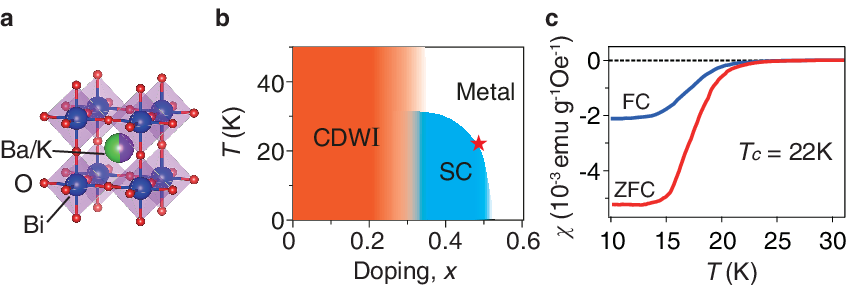}
\caption{\textbf{Crystal structure and phase diagram of Ba$_{1-x}$K$_x$BiO$_3$.} \textbf{a}, The perovskite-like crystal structure of Ba$_{1-x}$K$_x$BiO$_3$. % with space group $P$m3m.
The purple shades illustrate the BiO$_6$ octahedra. \textbf{b}, Phase diagram of Ba$_{1-x}$K$_x$BiO$_3$ according to Refs.~\onlinecite{CDW2,phasediagram}. The abbreviations stand for charge density wave insulator (CDWI) and superconductor (SC). The red marker shows the doping and $T_c$ of our samples. \textbf{c}, Temperature dependence of the zero field-cooled (ZFC) and field-cooled (FC) magnetic susceptibility of our Ba$_{0.51}$K$_{0.49}$BiO$_3$ single crystal measured at a magnetic field of 100~Oe. The sample shows a diamagnetic response and $T_c$ = 22~K.
}
\label{sample}
\end{figure}

The phase diagram of bismuthate superconductors fits the paradigm of high-temperature superconductivity emerging near a competing broken-symmetry phase.
As shown in Figs.~\ref{sample}a-b, BaBiO$_3$ is a perovskite-like insulator with a $\sim$2~eV band gap \cite{Optical_sato}, where a commensurate charge density wave (CDW) doubles the unit cell and is accompanied by breathing and tilting distortions of the BiO$_6$ octahedra \cite{CDW,CDW2,O_negative_DFT}.
With potassium doping, the CDW is suppressed and superconductivity develops, reaching its highest $T_c$ of 32~K  at Ba$_{1-x}$K$_x$BiO$_3$ with $x\sim 0.35$ (Ref.~\onlinecite{Sleight_review}).
%Bismuthate superconductivity exhibits unique characteristics setting it apart from both conventional superconductivity and other unconventional high-temperature superconductivity. 
These $T_c$s are much higher than those of conventional phonon-mediated superconductors with a similar density of states at the Fermi energy \cite{isotope2,Specific_heat_Hundley,Sleight_review}. However, unlike cuprates or iron-based superconductors where superconductivity emerges near magnetic order, there is no magnetic phase anywhere in the phase diagrams of the bismuthate superconductors, suggesting a nonmagnetic pairing mechanism \cite{Sleight_review,isotope2}. 

Two pictures have emerged to explain the bismuthates' long-debated pairing mechanism. One, rooted in the chemistry of the compound, posits that Cooper pairs form locally. The nominal Bi$^{4+}$ valence is not energetically favorable and charge disproportionates into Bi$^{3+}$ and Bi$^{5+}$ (Refs.~\onlinecite{Bi3+_5+,valence_Varma}), although the actual charge difference is much smaller than assumed in an ionic picture \cite{ZXShen_PRB,Bi4+}. Electron pairs preferentially occupying one sublattice could give rise to the insulating CDW state. Upon increasing doping, the charge disproportionation weakens and eventually disappears, but the electron pairs may survive and Bose condense, leading to superconductivity \cite{negativeU_Taraphder,valence_Varma}.
 It has also been stressed that one should think of these as negative charge-transfer materials, in which the holes introduced upon doping reside mainly in oxygen states of $A_{1g}$ symmetry around the Bi sites \cite{O_negative,O_negative2,CDW2}.

An alternative picture posits that the pairing in the bismuthates is due to strong electron-phonon coupling, with a model of electrons coupled to optical phonons introduced for BaPb$_x$Bi$_{1-x}$O$_3$ \cite{Rice_BaPb}. Early density functional theory (DFT) calculations, however, found an electron-phonon coupling (EPC) constant $\lambda$ of 0.34, far too small to account for the high $T_c$ (Ref.~\onlinecite{DFT_lambda}). More recently, it has been argued that the \textit{long range} Coulomb interactions (LRCI) were underestimated in bismuthates, since they are nearly insulators and the screening should be weak. Consequently,
the electron-phonon coupling is underestimated by DFT coupled with semi-local exchange-correlation functionals such as local density approximation (LDA) and generalized gradient approximation (GGA) \cite{Yinzhiping}. When LRCI exchange terms are considered  using the screened hybrid functionals and the GW method,
 the calculated bandwidth is significantly broadened, and electron-phonon coupling is increased substantially to $\lambda \simeq 1$, which could account for the high $T_c$ of $\sim$30~K (Ref. \onlinecite{Yinzhiping}).  
Furthermore, with such a large $\lambda$, dynamical mean field theory (DMFT) calculations reproduce the main features of optical conductivity studies \cite{DFT_optical_2012}. However, the band expansion effect of LRCI has never been directly corroborated in any real material \cite{ashcroft}.

To clarify the superconducting mechanism of this important superconductor family, it is crucial to obtain a comprehensive understanding of its  electronic structure and superconducting gap structure, which remain unknown. 
As one of the most direct probes of these properties, angle-resolved photoemission spectroscopy (ARPES) studies on bismuthate superconductors are still lacking,  possibly due to the difficult-to-cleave three-dimensional crystal structure and insufficient sample quality. 
Here we perform ARPES measurements on the electronic structure of high quality Ba$_{0.51}$K$_{0.49}$BiO$_3$ single crystals with a $T_c$ of 22~K (Fig.~\ref{sample}c), which is consistent with the reported phase diagram (Fig.~\ref{sample}b). %The K doping level is determined by electron probe micro-analysis (EPMA). X-ray diffraction measurements confirm the crystal structure (Supplementary Fig. S1), while 
%It has a simple cubic structure at room temperature \cite{inelastic_neutron,Sleight_review,Inelastic_neutron_2000} with $T_c$ of~22K, as shown in Figs.~\ref{sample}b-c. With decreased temperature, it goes through a cubic-to-tetragonal phase transition with merely 0.1\% increasing of the lattice constant $c$ \cite{Inelastic_neutron_2000}. A simple-cubic Brillouin zone is used for simplicity (Fig.~\ref{band}(a)). 

\begin{figure*}[t]
\centering
\includegraphics[width=140mm]{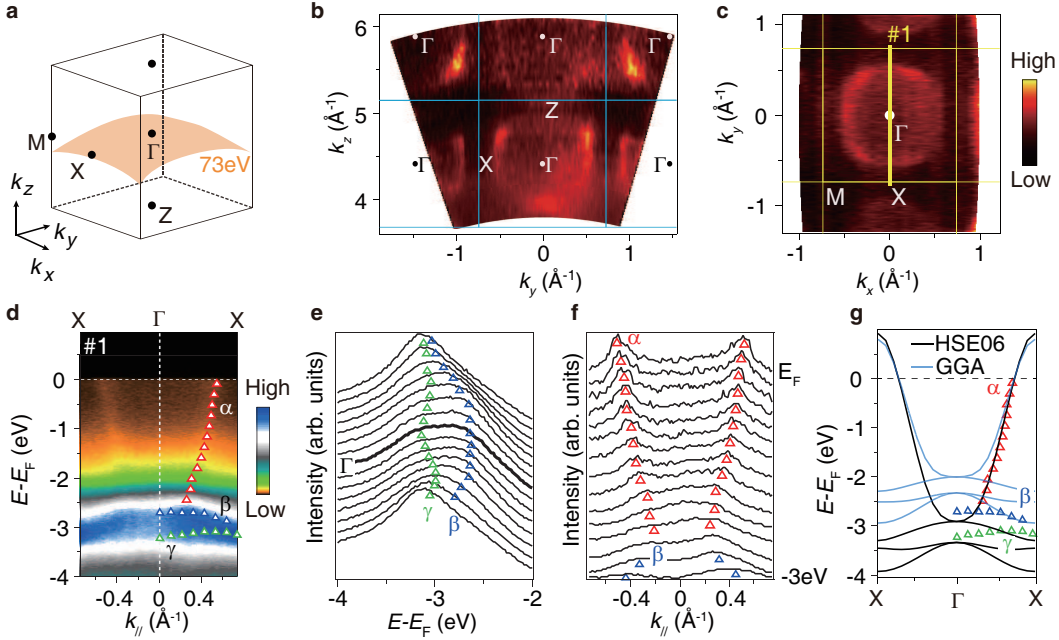}
\caption{\textbf{Electronic structure of Ba$_{0.51}$K$_{0.49}$BiO$_3$.} \textbf{a}, The Brillouin zone of Ba$_{0.51}$K$_{0.49}$BiO$_3$. The orange surface illustrates the momentum space sampled with $h\nu$ =73~eV photons near the $\Gamma$MX  plane. \textbf{b}, Photoemission intensity map in the $\Gamma$ZX plane integrated over an energy window of $E_F\pm$15~meV, which is measured with photons ranging from 57 to 132~eV. \textbf{c}, In-plane photoemission intensity map measured with $h\nu$ =73~eV photons, integrated over an energy window of $E_F\pm$15~meV. \textbf{d}, Photoemission intensity along cut $\#$1 shown in panel c. The bands $\alpha$, $\beta$, and $\gamma$ are indicated on the right side by markers in red, blue, and green, respectively. \textbf{e}, Energy distribution curves (EDCs) of data in the lower part of panel d, showing the dispersions of bands $\beta$ and $\gamma$. \textbf{f}, Momentum distribution curves (MDCs) of data in the upper part of panel d, showing the dispersions of bands $\alpha$ and $\beta$. \textbf{g}, Band dispersion extracted from the data in panel d (markers in red, blue, and green) plotted over the density functional theory (DFT) calculations of BaKBi$_2$O$_6$ using generalized gradient approximation (GGA, blue curves) and Heyd-Scuseria-Ernzerhof hybrid functional calculations (HSE06, black curves). 
}
\label{band}
\end{figure*}

The three-dimensional Fermi surface structure is revealed by the combination of $k_z$ dependent and in-plane photoemission intensity maps at the Fermi energy (Figs.~\ref{band}a-c). The Fermi surface cross-sections in the $\Gamma$ZX plane match the period of the Brillouin zones when assuming an inner potential of 7~eV (Fig.~\ref{band}b) and match the cross-section in the $\Gamma$MX plane (Fig.~\ref{band}c) as expected. The Fermi surface is a rounded cube shape centered at $\Gamma$, consistent with theoretical calculations \cite{DFT_lambda,FS_cal_2000,FS_Compton}. 
Based on the Fermi surface volume and Luttinger's theorem \cite{Luttinger_2}, the electron carrier density is estimated to be 0.48$\pm0.05$~e$^{-}$/unit cell. The photoemission intensity along $\Gamma$--X shows three bands near the Fermi energy (Fig.~\ref{band}d). The energy distribution curves (EDCs, Fig.~\ref{band}e) show two flat bands ($\beta$ and $\gamma$) around 3~eV below the Fermi energy, while the momentum distribution curves (MDCs) show an electron-like band ($\alpha$) crossing the Fermi level (Fig.~\ref{band}f) to form the electron-like Fermi surface (Figs.~\ref{band}b-c).

% PRL style: We resolved three-dimensional Fermi surface and detailed band structure, which is almost identical to the GW calculations and DFT-HSE06-a poor man's GW-calculations, but differs largely from the DFT-GGA calculations. This suggests that for materials in the vicinity of an insulating phase, it is critical to include long-range exchange interactions in the first-principles calculations, in order to overcome the over-screening effects of GGA or LDA. Moreover, we observed isotropic superconducting gap on the Fermi surface, and a kink structure at 50~meV in the electronic band dispersion, which gives a electron-phonon coupling constant of 1.2. This value is much larger than the DFT-LDA calculated value and is consistent with the GW/HSE06 calculations, and can account for the high $T_c$ in Ba$_{0.51}$K$_{0.49}$BiO$_3$. These suggest that Ba$_{0.51}$K$_{0.49}$BiO$_3$ is an s-wave superconductor in the intermediate coupling regime.  Our work establishes  Ba$_{1-x}$K$_{x}$BiO$_3$ as a benchmark BCS oxide superconductor, whose superconductivity originates from long-range exchange interaction driven strong electron-phonon coupling, and sheds new lights on finding new high-Tc superconductors.

%To reveal the electron correlation strength in Ba$_{0.51}$K$_{0.49}$BiO$_3$, t
%Compared with the calculated band structure using GGA (Fig.~\ref{band}g), 
At $\Gamma$,   $\alpha$ and $\beta$  are degenerate as shown by the calculations (Fig.~\ref{band}g). The measured occupied bandwidth of $\alpha$ is unexpectedly $\sim$34\,\%
larger than that calculated using GGA, in stark contrast to the cuprate and iron-based superconductors, where the electronic bands are strongly renormalized to be much {\it narrower} than the DFT-LDA/GGA bandwidth due to short-range Coulomb interactions. Our findings indicate that the short-range Coulomb interaction strength is very weak in Ba$_{0.51}$K$_{0.49}$BiO$_3$, fundamentally different from in cuprate and iron-based superconductors.
On the other hand, the highly-dispersing $\alpha$ band is reproduced excellently by DFT-HSE06 calculations (using Heyd-Scuseria-Ernzerhof hybrid functional, Fig.~\ref{band}g), without any tuning parameter. 
%Although the broad photoemission feature of band $\beta$ leads to uncertainty of the $\alpha$ band bottom, the discrepancy in occupied bandwidth is within $\sim$8\,\%. 
A slight discrepancy near the $\alpha$  band bottom leads to a $\sim$8\,\% narrower occupied bandwidth than in the HSE06 calculation, perhaps arising from uncertainty due to broad photoemission features, or finite but weak short-range Coulomb interactions. 
The HSE06 hybrid functional (and GW method) includes the
%screened 
long-range exchange interaction, which results from LRCI and is largely omitted in the semi-local LDA and GGA functionals. 
This exchange interaction tends to delocalize conduction electrons and increase their kinetic energy and bandwidth \cite{ashcroft}.
LRCI often plays an important role in systems with low carrier density or poor screening, such as Mott insulators \cite{LRC1}, semiconductors  \cite{LRC2}, and semimetals \cite{LRC3}, and the near-perfect
 reproduction of the observed bandwidth by HSE06 here indicates that it plays an important role in the bismuthate superconductors, likely because bismuthates are near an insulating phase. 
 As discussed later, this interaction not only expands the electronic bands, but also has a profound impact on the lattice dynamics and electron-phonon interaction.
  %To be done, add formula and discussions on how electronic correlation (self-energy) impact the electron-phonon coupling, reference to the APS talk.
  %More importantly, \textbf{(zhiping add more discussions on the band comparison)}

\begin{figure}[t]
\centering
\includegraphics[width=100mm]{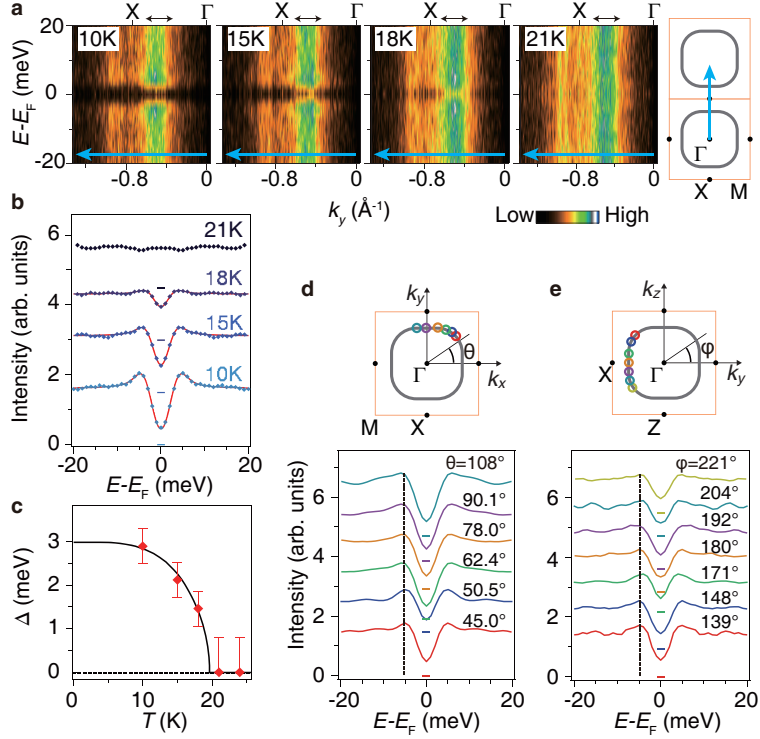}
\caption{\textbf{Temperature and momentum dependence of the superconducting gap.} \textbf{a}, Temperature dependence of symmetrized photoemission spectra along the $\Gamma$X direction, measured in the $\Gamma$MX plane by 30~eV photons. The right inset illustrates the Brillouin zones (orange squares), Fermi surfaces (gray rounded squares), and the photoemission cut (blue arrow). \textbf{b}, Temperature dependence of the symmetrized EDCs (diamond shaped markers) integrated around $k_F$ (double arrows in panel a). The EDCs are vertically offset for better visualization, while the horizontal bars indicate the zero position of each EDC. The red solid curves are the fitting with resolution-convolved Dynes functions. \textbf{c}, Temperature dependence of the superconducting gap (red markers) fit to the BCS function (solid black curve). The error bars are determined by combining the standard deviation of the fitting by Dynes functions (panel c) and the energy uncertainty. \textbf{d}, Symmetrized EDCs at various $k_F$'s (as shown in the top panel with the one-to-one corresponding colors) of band $\alpha$ in the $\Gamma$MX plane measured at 10~K. The EDCs are vertically offset for better visualization, while the horizontal bars indicate the zero positions of each EDC. The top panel illustrates the Brillouin zone (orange square), Fermi surface cross-section (gray rounded square), and the color-coded Fermi momenta   (open circles). \textbf{e}, Same as panel d but along the $k_F$'s in the $\Gamma$ZX plane.}
\label{T-dep}
\end{figure}

The temperature dependence of the superconducting gap is investigated at the Fermi momentum ($k_F$) along $\Gamma$X in the $\Gamma$MX plane using 30~eV photons (Fig.~\ref{T-dep}a).
%shows the temperature dependence of symmetrized intensity distribution taken by using $h\nu$ = 30~eV along $\Gamma-X$ direction, around the Fermi momentum ($k_F$).  
At 10~K, the symmetrized photoemission intensity is suppressed at $E_F$, indicating the opening of an energy gap. Upon increasing temperature, the gap gradually decreases and finally closes at 21~K, consistent with the $T_c$ measured by magnetic susceptibility (Fig.~\ref{sample}c). Below $T_c$, the symmetrized EDCs integrated around $k_F$ show a superconducting gap with a coherence peak (Fig.~\ref{T-dep}b). We fit the EDCs to the Dynes function \cite{Dynes_function}, $N(E, \Gamma)=|{Re[(E-i\Gamma)/\sqrt{(E-i\Gamma)^2-\Delta^2}]}|$, where $N(E,\Gamma)$ is the measured spectrum, $\Delta$ is the gap and $\Gamma$ is a broadening term (also called the scattering rate). After convolving the energy resolution, we get $\Delta$(10~K) = 2.9~meV and $\Gamma$(10~K) = 0.04~meV. Upon increasing temperature, the coherence peak intensity decreases and the gap closes. The temperature dependence of the gap fits well to the BCS formula, giving a 2$\Delta$(0)/$k_BT_c$ of 3.51 $\pm$ 0.05, consistent with previous reports on overdoped samples \cite{photoemission}.
%close to the previous findings of 2$\Delta$(0)/$k_BT_c$$\sim$3.9 for Ba$_{0.54}$K$_{0.46}$BiO$_3$ by angle-integrated photoemission spectroscopy \cite{photoemission}, indicating that the superconductivity is in an intermediate coupling regime.
The superconducting gap structure in momentum space, which reflects the pairing symmetry, was investigated  
along the Fermi surface cross-sections in the high-symmetry $\Gamma$MX and $\Gamma$ZX planes (Figs.~\ref{T-dep}d and e). 
As shown by the dashed lines tracking the coherence peaks, the gap remains constant within our uncertainty, indicating an isotropic gap.

% In their measurement, a pseudogap located at 70meV is observed, which is claimed to originate from a strong electron-phonon coupling. Considering the different doping level of the samples, we propose that the coupling strength may depend on the K doping amount.

\begin{figure}[t]
\centering
\includegraphics[width=86mm]{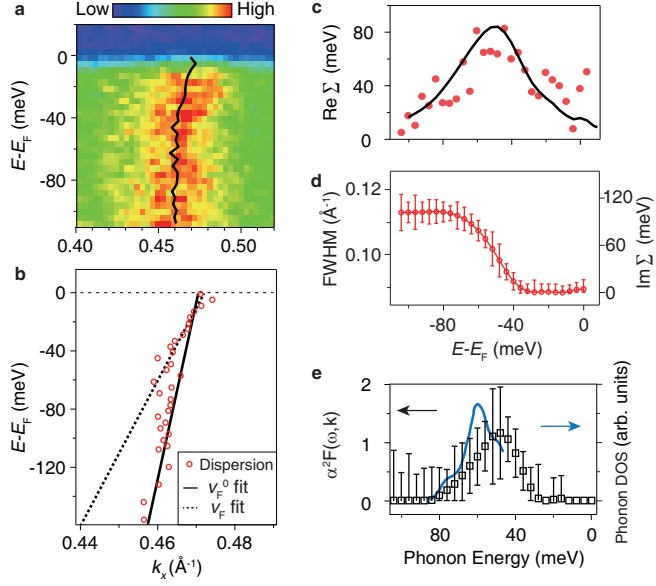}
\caption{\textbf{Electron-phonon interaction in Ba$_{0.51}$K$_{0.49}$BiO$_3$.} \textbf{a}, ARPES spectrum taken along $\Gamma$X using  72~eV photons. The band dispersion from Lorentz fitting on the MDCs is overlaid. \textbf{b}, MDC-derived dispersion from the raw data (open circles), and linear fits to the high energy ($E-E_F<$-80meV, $v_F^0$ fit, solid line) and low energy ($E-E_F>$-50meV,  $v_F$ fit, dashed line) dispersions. \textbf{c}, Real part of the self-energy Re$\Sigma$ obtained in two ways: the difference between the band dispersion near $E_F$ and the polynomial fit approximating the bare band dispersion (red dots), and the Kramers-Kronig transformation of the imaginary part of the self-energy Im$\Sigma$ (black curve). \textbf{d}, Full-width at half-maximum (FWHM) of the MDCs, and Im$\Sigma$ computed from the FWHM. \textbf{e}, Electron-phonon spectral function $\alpha^2F(\omega, k)$ (black squares) calculated from the Im$\Sigma(\omega)$. To avoid unphysical values, we have set all negative data points  of $\alpha^2F(\omega, k)$ to zero. The blue curve is the phonon density of states (DOS) determined by neutron diffraction in Ref.~\onlinecite{Inelastic_neutron_Loong}.  }
\label{kink}
\end{figure}

Upon closer inspection of the ARPES spectra near $k_F$, a kink in the  dispersion can be observed around a binding energy of 50~meV (Fig.~\ref{kink}a), which is the signature of electron-boson interactions \cite{Eliashberg_Hofmann}. In the dispersion extracted from the Lorentz fitting on the MDCs (Fig.~\ref{kink}b), the slope between $E_F-50$~meV and $E_F$ gives the  Fermi velocity, $v_F$, 
and the slope on a larger energy scale approximates the bare-band Fermi velocity, $v_F^0$. Their difference indicates a strongly enhanced band mass.
The difference between the low energy dispersion and the polynomial fit of the bare band  gives the real part of the self-energy, Re$\Sigma$ (Fig.~\ref{kink}c), which shows a maximum around 50~meV.
 The full-width at half-maximum (FWHM) of the MDCs also exhibits a prominent increase around 50~meV (Fig.~\ref{kink}d), indicating a major change in the quasiparticle lifetime. 
The quasiparticle lifetime at $E_F$ is not infinite as expected for an ideal Fermi liquid, since the FWHM is finite at $E_F$ due to extrinsic experimental angular broadening. To exclude this, we subtract the minimum value of the FWHM around $E_F$ as a constant background; then the FWHM is multiplied by  $v_F^0$ to obtain the imaginary part of the self-energy, Im$\Sigma$.
%Since the short-range Coulomb interaction is small in this system (Fig.~2), the Fermi liquid model should apply and the quasiparticle lifetime should be infinite at the $E_F$. %Thus the finite FWHM of MDCs at $E_F$ is from 
%Therefore, the non-zero FWHM of MDCs at $E_F$ is due to the extrinsic momentum broadening by experimental setup, which can be considered as a constant background. 
%To estimate the imaginary part of the self energy, Im$\Sigma$, we subtract the FWHM of the MDCs by its minimum value around $E_F$ as a constant background, and then multiply it by the Fermi velocity of bare band dispersion (Fig.~\ref{kink}d). %Im$\Sigma$ shows a prominent increase in the range of $-50\pm$20~meV (Fig.~\ref{kink}d), indicating a major change in the quasiparticle lifetime due to electron-boson interactions. 
%is obtained by fitting the width of the MDCs and multiply it with the Fermi velocity of bare band dispersion (Fig.~\ref{kink}d). A prominent increase of Im$\Sigma$ is observed in the range of -50$\pm$20~meV.
The Re$\Sigma$ from the Kramers-Kronig transformation of Im$\Sigma$ matches well with that obtained from the dispersion (Fig.~\ref{kink}c), indicating that Re$\Sigma$ and Im$\Sigma$ are self-consistent (Supplementary Fig. S2). %In Fig.~\ref{kink}(c), the experimental Re$\Sigma$ matches well to that computed from KK transform of the Im$\Sigma$ data in Fig.~\ref{kink}(d). This indicates that our self-energy analysis is self-consistent.
 The kink  and the abrupt change of Im$\Sigma$ within $\left[-35~meV,-70~meV\right]$ demonstrate strong interactions between electrons and bosonic modes, and coincide with an oxygen-breathing mode found around 60~meV by neutron scattering (Fig.~\ref{kink}e) \cite{Inelastic_neutron_Loong} and predicted to couple with electrons most strongly \cite{Yinzhiping}. 
Therefore, the observed  self-energy behavior is most likely due to strong coupling of the conduction band to the oxygen-breathing phonon.

The electron-phonon coupling strength, $\lambda$, is central to the controversy around the superconducting mechanism of Ba$_{1-x}$K$_{x}$BiO$_3$.  In a simple estimate $\lambda=v_F^0/v_F-1$ = 1.4. Alternatively, one can compute $\lambda$ from the self-energy. According to the Migdal-Eliashberg formalism, the electron-phonon spectral function, $\alpha^2F(\omega, k)$, is obtained by
%related to the imaginary part of the self-energy by Im$\Sigma(\omega)=\pi\int_{0}^{\omega_{max}}\alpha^2F(\omega)$\,d$\omega$. Thus 
the derivative of Im$\Sigma(\omega)$, % gives $\alpha^2F(\omega, k)$, 
which is in good agreement with the phonon DOS (Fig.~\ref{kink}e). 
Then the coupling strength $\lambda$ is calculated to be 1.3 $\pm$  0.2 (Ref.~\onlinecite{Eliashberg_Hofmann}). %obtained by the integration $\lambda = 2\int_{0}^{\omega_{max}}\alpha^2F(\omega, k)/\omega$\,d$\omega$ = 1.3 $\pm$  0.2 (Ref.~\onlinecite{Eliashberg_Hofmann}) (Fig.~\ref{kink}e). 
%Therefore, we estimate that  $\lambda \approx 1.3 \pm 0.2$ in Ba$_{0.51}$K$_{0.49}$BiO$_3$. 
Using the McMillan equation, $T_c=(\Theta_D/1.45)exp\{-1.04(1+\lambda)/[\lambda-\mu^*(1+0.62\lambda)]\}$ with the Debye temperature $\Theta_D=280$~K (Ref.~\onlinecite{Specific_heat_Hundley}) and  $\lambda =$ 1.3, we get $T_c$ = 22~K with a Coulomb pseudopotential parameter $\mu^*$ = 0.11, which is within the usual assigned range of 0.1$\sim$0.15.
Thus the large  $\lambda$ is sufficient to explain the high $T_c$ of this bismuthate superconductor.
Based on this analysis and the observed self-energy behavior and isotropic gap structure, we conclude that Ba$_{0.51}$K$_{0.49}$BiO$_3$ is a BCS superconductor.

Revealing the electron-phonon coupling strength is essential for a quantitative understanding of both the conventional BCS superconductors 
%such as MgB$_2$ 
\cite{MgB2_review,H2S}, 
% and H$_3$S under pressure 
 and even some unconventional superconductors where electron-phonon coupling plays a substantial role, such as monolayer FeSe/SrTiO$_3$ \cite{ZXNature,Songqi}. When the electronic self-energy $\Sigma(k,\omega)$ is negligible, linear response calculations based on LDA/GGA can compute the electron-phonon coupling strength well.
However, a large LRCI, as in Ba$_{0.51}$K$_{0.49}$BiO$_3$, leads to a large momentum-dependent self-energy and an expansion of the electronic band $\varepsilon(k)\approx(1+a)\varepsilon_{LDA}(k)+b$. As a result, the electron-phonon dynamical matrix element is roughly $(1+a)$ times the LDA/GGA value and the mode-dependent electron-phonon coupling strength is roughly $(1+a)^2$ times stronger. In Ba$_{0.51}$K$_{0.49}$BiO$_3$, the GGA conduction bandwidth is $\sim$2.8 eV whereas the HSE06 bandwidth is $\sim$3.9 eV, resulting in $a=0.4$ and a factor of $\sim$2 enhancement of the GGA electron-phonon coupling strength.
Furthermore, the LRCI tends to delocalize electrons and overcome the over-binding problem of LDA and GGA, which gives the correct band gap for the insulating parent compound BaBiO$_3$ and a further enhancement of the LDA/GGA electron-phonon coupling. It was found that the combined effects due to the inclusion of LRCI result in  the average coupling strength  between electrons and oxygen breathing and stretching  phonons to be $\sim$1.0, a factor of $\sim$3 enhancement over the LDA electron-phonon coupling strength \cite{Yinzhiping}. The agreement between this calculation and our experimental findings   highlights the profound impact of the LRCI on electron-phonon coupling.
% which gives $\lambda \approx 1.3 \pm 0.2$  contributed mainly by the oxygen-breathing mode around 60~meV. 
%\sout{As  shown in ref. \onlinecite{Yinzhiping}, such a large  $\lambda$ could fully account for the high $T_c$ of the bismuthate superconductors, using the modified McMillan equation \cite{Dynes1}.

%{\color{blue}
The observed large bandwidth that can only be correctly reproduced by including long-range exchange interactions constitutes the first demonstration of significant bandwidth expansion introduced by the Fock term, \textit{i.e.} LRCI, in the Hartree-Fock treatment of a many-body system. The exchange-hole effects reduce the electronic energy unequally  in momentum space and thus expand the bands, as elaborated in many textbooks, for example in Fig.~17.1 of Ashcroft \& Mermin's \textit{Solid State Physics} \cite{ashcroft}. 
In real materials, there are always two opposite effects with comparative orders of magnitude --- 
the frequency dependence of self-energy due to short-range Coulomb interactions narrows the bands, while the momentum dependence of self-energy arising from LRCI expands them.  For metals with $s$ and $p$ orbitals, the bandwidth calculated by LDA or GGA is usually within 10\% of the experimental value, making it difficult to separate these two effects. In Ba$_{0.51}$K$_{0.49}$BiO$_3$,
  the narrowing effect is small, while the expansion effect is huge, which 
provides a unique opportunity to uncover the LRCI band expansion.  Our data and calculations represent the first direct demonstration of this fundamental effect.

To summarize, we here present the first comprehensive characterization of the electronic structure of a bismuthate superconductor, showing that Ba$_{0.51}$K$_{0.49}$BiO$_3$ is a benchmark BCS superconductor with an isotropic superconducting gap and 2$\Delta$/$k_B$$T_c$ = 3.51 $\pm 0.05$. The sizable electron-phonon coupling strength ($\lambda \approx 1.3$)   can account for the high $T_c$, solving a 30-year mystery. 
Moreover, the remarkable agreement between our data and the screened hybrid functional calculations represents the first direct experimental proof that including 
long-range Coulomb interactions is crucial for compounds with low carrier density or poor screening, even if the on-site Coulomb interactions are weak, 
so that one can overcome the over-binding problem of LDA or GGA and obtain the correct band structure and electron-phonon coupling strength. This is particularly critical for the reliable prediction of new superconductors.

\vskip 4mm

\noindent \textbf{Methods}

\noindent \textbf{Sample synthesis and characterization:} High quality Ba$_{0.51}$K$_{0.49}$BiO$_3$ single crystals were synthesized by the self-flux method as described elsewhere \cite{HHWen}. The chemical composition was determined by electron probe micro-analysis (EPMA) and normalized to the stoichiometric value of Bi. The crystal structure was verified by x-ray diffraction (Supplementary Fig. S1). The superconductivity was confirmed through magnetic susceptibility measurements using a Quantum Design SQUID VSM.

\noindent \textbf{ARPES measurements:}  The electronic structure in Fig.~\ref{band} was measured at Shanghai Synchrotron Radiation Facility beamline 09U with a Scienta DA30 analyzer, and the energy resolution was 18~meV. The superconducting gap was measured at Diamond Light Source beamline I05 with a Scienta R4000 analyzer, and the energy resolution was  better than 4~meV. The data in Fig.~\ref{kink} were taken at Advanced Light Source (ALS) beamline 4.0.3 with a Scienta R8000 analyzer, and the energy resolution was 12~meV. 
All samples were cleaved and measured under a vacuum better than 5$\times$10$^{-11}$ mBar. 

\noindent \textbf{ARPES spectra analysis:} 
According to the Migdal-Eliashberg formalism, the electron-phonon spectral function, $\alpha^2F(\omega, k)$, is related to the imaginary part of the self-energy by Im$\Sigma(\omega)=\pi\int_{0}^{\omega_{max}}\alpha^2F(\omega)$\,d$\omega$. 
Thus we calculate $\alpha^2F(\omega, k)$ by the derivative of Im$\Sigma(\omega)$. 
Then the coupling strength is calculated according to the formula $\lambda = 2\int_{0}^{\omega_{max}}\alpha^2F(\omega, k)/\omega$\,d$\omega$ (Ref.~\onlinecite{Eliashberg_Hofmann}).

\noindent \textbf{Band structure calculations:}  Band structures were calculated using the VASP package \cite{vasp} with GGA (in the Perdew-Burke-Ernzerhof form \cite{PBE}) and HSE06 exchange-correlation functionals \cite{HSE06,HSE_2011}. We used the simple-cubic perovskite structure with lattice constant $a$ = 4.27\,\AA, an energy cutoff of 500\,eV and a $20\times20\times20$ $k$ mesh in both DFT-GGA and DFT-HSE06 calculations.

\vskip 4mm
\noindent \textbf{Acknowledgements}\\
We gratefully acknowledge enlightening discussions with Prof. Z.-X. Shen, Prof. C. M. Varma, Prof. G. A. Sawatzky and Dr. D. Peets, and the experimental support of Dr. J. Denlinger, Dr. Q. Q. Ge, Dr. Y. B. Huang, Dr. Z. H. Chen, Dr. T. Kim., and  Dr. Z. Sun,
We thank the Diamond Light Source for  time on beam line I05 under Proposal No. SI11914, the Shanghai Synchrotron Radiation Facility for access to beamline 9U, and the Advanced Light Source for access to beamline 4.0.3. Some preliminary data were taken at  National Synchrotron Radiation Laboratory (NSRL, China).
This work is supported by the National Key R\&D Program of the MOST of China (Grants Nos.\ 2016YFA0300200, 2017YFA0303004, 2016YFA0302300, and 2016YFA0300400), the National Natural Science Foundation of China (Grants Nos.\ 11574337, 11227902, U1332209, 11704073, 11504342, 11674030, and 11534005), and the Fundamental Research Funds for the Central Universities (Grant No.\ 310421113). The calculations used high performance computing cluster of Beijing Normal University in Zhuhai.

\vskip 4mm
\noindent \textbf{Author contributions}\\
Single crystal samples were grown by Y. J. J., T. F. D., and H. H. W..
ARPES measurements were performed by C. H. P. W., R. P., Q. Y., X. H. N., Q. Y. C., Z. T. L., D. W. S. and Q. S..
Sample characterization was performed by C. H. P. W., Q. Y., X. L., Y. F. F., X. S. L., and Y. H. S.
The data analysis was performed by C. H. P. W., H. C. X., R. P., and D. L. F..
Z. P. Y. performed the calculations, Z. P. Y. and G. K. contributed to the theoretical interpretation of the experiments. 
 D. L. F., H. C. X., R. P., Z. P. Y., C. H. P. W. and Q. Y. wrote the manuscript. D. L. F. and H. H. W. coordinated the project. 
All authors have discussed the results and the interpretation.

\vskip 4mm
\noindent \textbf{Additional information}\\
Supplementary information is available in the online version of the paper.

\newpage

\end{document}